\documentclass[10pt,twocolumn]{article}

\newcommand{\comment}[1]{}
\newcommand{\keywords}[1]{}

\begin{document}

\title{Rotating Stars in Relativity}

\author{Nikolaos Stergioulas\\
\\
Department of Physics, University of Wisconsin-Milwaukee\\
PO Box 413, Milwaukee, WI 53201, USA\\
and\\
Max Planck Institute for Gravitational Physics\\ 
(Albert-Einstein-Insitute)\\
D-14473 Potsdam, Germany\\
email: niksterg@aei-potsdam.mpg.de\\
\\
Review article for {\em Living Reviews in Relativity}\\
{\tt http://www.livingreviews.org} }
\date{}
\maketitle


{\bf Abstract.}
Rotating relativistic stars are receiving significant attention in recent
years, because of the information they can yield about the equation of state
of matter at extremely high densities and because they are one of the more
possible sources of detectable gravitational waves. We review the latest
theoretical and numerical methods for modeling rotating relativistic stars,
including stars with a strong magnetic field and hot proto-neutron stars.
We also review nonaxisymmetric oscillations and instabilities in rotating
stars and summarize the latest developments regarding the gravitational
wave-driven (CFS) instability in both polar and axial quasi-normal modes.


\section{Introduction}

Rotating relativistic stars are of fundamental interest in physics. Their
bulk properties restrict the proposed possible equations of state for
densities larger than nuclear density. Their oscillations can become
unstable, producing gravitational waves that could be detectable, providing 
thereby a new way of probing the interior of neutron stars.

Recent research has considerably advanced our understanding of these
objects. There now exist several independent numerical codes for
obtaining accurate models of rotating neutron stars in full general
relativity. Three of these codes have been shown to agree with each other to
remarkable accuracy and one code is available as public domain for use by
other researchers.

The numerically constructed maximum mass models, for different proposed
equations of state, differ by as much as a factor of two in mass, radius and
angular velocity, a factor of five in central density and a factor of eight
in the moment of inertia. These large uncertainties show that our
understanding of the properties of matter at very high densities is currently
rather poor.

Despite the different maximum rotation rates, corresponding to 
different candidates for the equation of state of neutron-star matter, 
one can place an absolute upper limit on the rotation of
relativistic stars, by imposing causality as the only
requirement on the equation of state. It then follows that gravitationally
bound stars cannot rotate faster than 0.28 ms.

Although observed magnetic fields in neutron stars have a negligible
effect on neutron-star structure, a sufficiently  strong magnetic field acts as a 
centrifugal force on a relativistic star,
flattening its shape and increasing the maximum mass and rotation rate for
a given equation of state. The magnetic field strength of a stationary
configuration has been shown to have an upper limit of $B \sim 10^{17}$ G.

Rapidly rotating proto-neutron stars are shown to have an extended envelope,
due to their high temperature and the presence of trapped neutrinos. If the
equation of state is softened, as the neutron star cools, by a large
amplitude phase transition, then the nascent neutron star may collapse to a
black hole. A surprising result is that a supramassive proto-neutron star,
even though it contracts during cooling, evolves to a cold neutron star of
smaller angular velocity.

In rotating stars, nonaxisymmetric perturbations have been studied in the
Newtonian and post-Newtonian approximations, in the slow-rotation limit and
in the Cowling approximation but fully relativistic quasi-normal modes (except 
for
neutral modes) have yet to be obtained. The effect of rotation on the 
quasi-normal
modes of oscillation is to couple polar and axial modes and to shift
their frequencies and damping times causing some modes to become unstable.

Nonaxisymmetric instabilities in rotating stars can be driven by the emission
of gravitational waves (CFS-instability) or by viscosity. The onset of the
CFS-instability has now been computed for fully relativistic, rapidly
rotating stars.  Relativity has a strong influence on the onset of the
instability, allowing it to occur for less rapidly rotating stars than was
suggested by Newtonian computations.

Contrary to what was previously thought, nascent neutron stars can be
subject to the $l=2$ bar mode CFS-instability, emitting strong gravitational
waves. The frequency of the waves sweeps downward through the optimal 
LIGO sensitivity window and first estimates show that it could be detectable 
out to the distance of 140 Mpc by the advanced LIGO detector.

The viscosity-driven instability is not favored by general relativity but,
as a new relativistic computation shows, is absent in rotating
neutron stars, unless the equation of state is unexpectedly stiff.

Axial fluid modes in rotating stars ($r$-modes) received renewed attention
since it was discovered that they are generically unstable to the emission of
gravitational waves.  The $r$-mode instability can slow down a newly-born
rapidly rotating neutron star to Crab-like rotation rates. First results show that,
the gravitational waves from the spin-down (directly, or as a stochastic
background) could be detectable by the advanced LIGO or VIRGO detectors.

The present article aims at presenting a summary of theoretical and numerical
methods that are used to describe the equilibrium properties of rotating
relativistic stars and their oscillations. It focuses on the most recently
available preprints, in order to rapidly communicate new methods and results.
At the end of some sections, the reader is pointed to papers that could not 
be presented in detail here. As new developments in the field occur, updated
versions of this article will appear.

\section{The Equilibrium Structure of Rotating Relativistic Stars}

\subsection{Assumptions}

     Although a relativistic star has a complicated structure (solid
crust, magnetic field, possible superfluid interior etc.), its bulk
properties can be computed with reasonable accuracy by making several
simplifying assumptions. 

The matter is modeled as a perfect fluid because observations of
pulsar glitches have shown that the departures from perfect fluid
equilibrium due to the solid crust are of order $10^{-5}$ \cite{FI92}.
The temperature of a cold neutron star is assumed to be 0 K because
its thermal energy ($<<1$MeV $\sim 10^{10}$ K) is much smaller than
the Fermi energy of the interior ($> 60$ MeV). One can then use a
zero-temperature (one-parameter) equation of state (EOS) to describe
the matter:
\begin{equation}
     \epsilon = \epsilon(P),
\end{equation}
where $\epsilon$ is the energy density and $P$ is the pressure. 
At birth, a neutron star is differentially rotating but as the neutron
star cools, shear viscosity, resulting from neutrino diffusion,
aided by convective and turbulent motions and possibly by the winding-up
of magnetic field lines, enforces uniform rotation. At present, it is
difficult to accurately compute the timescale in which uniform rotation
is enforced, but it is estimated to be of the order of seconds
\cite{GHZ98}. 

Within roughly a year after its formation, the neutron star temperature
becomes less than $10^9$K and its outer core becomes superfluid
(see \cite{Me98} and references therein). Rotation causes the 
superfluid neutrons to form an array of quantized vortices, with an
intervortex spacing of 
\begin{equation}
  d_n \sim 3.4 \times 10^{-3} \Omega_2^{-1/2} {\rm cm},
\end{equation}
where $\Omega_2$ is the angular velocity of the star in $10^2 {\rm s}^{-1}$.
On scales much larger than the intervortex spacing, e.g. on the order
of 1 cm, the fluid motions can be averaged and the rotation can
be considered  uniform \cite{So87}.The error in computing the metric is of order
\begin{equation}
     ( \frac{1 {\rm cm}}{R} )^2 \sim 10^{-11},
\end{equation}
where R is a typical neutron star radius \cite{FI92}. 

The above arguments show
that the bulk properties of a rotating relativistic star can be
modeled accurately by a uniformly rotating, zero-temperature perfect
fluid.

\subsection{Geometry of Space-Time}

     In relativity, the space-time geometry of a rotating star in
equilibrium is described by a stationary and axisymmetric metric of
the form
\begin{equation}
  ds^2 = -e^{2 \nu} dt^2 + e^{2 \psi} (d \phi - \omega dt)^2 + e^{2 \alpha}
            (dr^2+r^2 d \theta^2),    \label{e:metric}
\end{equation}
where $\nu$, $\psi$, $\omega$ and $\alpha$ are four metric functions which
depend on the coordinates $r$ and $\theta$ only (unless otherwise noted,
we assume $c=G=1$). The perfect fluid has a stress-energy
tensor
\begin{equation}
     T^{ab} = (\epsilon+P)u^a u^b + P g^{ab},
\end{equation}
a four velocity
\begin{equation}
     u^a = \frac{e^{-\nu}}{\sqrt{1-v^2}} (t^a + \Omega \phi^a),
\end{equation}
and a 3-velocity with respect to a zero angular momentum observer of
\begin{equation}
     v = (\Omega-\omega)e^{\psi-\nu},
\end{equation}
where $t^a$ and $\phi^a$ are the two killing vectors associated with the
time and translational symmetries of the space-time, $g_{ab}$ is the
metric tensor and $\Omega$ is the angular velocity. Having specified an
equation of state for very dense matter, the structure of the
star is computed by solving four components of 
Einstein's gravitational field equations
\begin{equation}
     R_{ab} = 8 \pi (T_{ab}- \frac{1}{2} g_{ab}T),
\end{equation}
(where $R_{ab}$ is the Ricci tensor and $T=T_a^a$) and the equation of  
hydrostationary equilibrium.

\subsection{Equations of State}

     The simplest equation of state one can use to model relativistic
stars is the relativistic polytropic EOS \cite{T65}
\begin{equation}
     P = K \rho^\Gamma,
\end{equation}
\begin{equation}
     \epsilon = \rho c^2 + \frac{P}{\Gamma-1},
\end{equation}
where $\rho$ is the rest mass density, $K$ is a constant and $\Gamma$ 
is the polytropic exponent. Instead of $\Gamma$, one often uses
the polytropic index $N$, defined through
\begin{equation}
  \Gamma=1+\frac{1}{N}.
\end{equation}
For this equation of state, the quantity $c^{(\Gamma-2)/(\Gamma-1)}
\sqrt{K^{1/(\Gamma-1)}/G}$ has units of length. In gravitational units
($c=G=1$), one can thus use $K^{N/2}$ as a fundamental length scale
to define dimensionless quantities.Equilibrium models are then
characterized by the polytropic index $N$ and their properties can be 
scaled to different values, using an appropriate value for $K$.
For $N<1.0$ ($N>1.0$) one
obtains stiff(soft) models, while for $N=0.5 - 1.0$, one obtains models with
bulk properties that are comparable to those of observed neutron stars.

Note that the for the above polytropic EOS, the polytropic index $\Gamma$
coincides with the adiabatic index of a relativistic isentropic fluid
\begin{equation}
  \Gamma = \frac{\epsilon+P}{P} \frac{dP}{d \epsilon}. \label{adiab}
\end{equation}
This is not the case for the polytropic equation of state
 $P= K \epsilon ^\Gamma$, that has been used
by other authors, which satisfies  (\ref{adiab}) only in the Newtonian
limit.

     The true equation of state that describes the interior of compact
stars is largely unknown. This results from the inability to verify
experimentally the different theories that describe the strong
interactions between baryons and the many-body theories of dense
matter at densities larger than about twice the nuclear density
(i.e. at densities larger than about $5\times 10^{14} {\rm gr}/{\rm
cm}^3$). 

Many different realistic EOSs have been proposed to date which all
produce neutron stars that satisfy the currently available
observational constraints (currently, the two main constraints are
that the EOS must admit nonrotating neutron stars with gravitational
mass of at least $1.44 M_{\odot}$ and allow rotational periods at
least as small as $1.56$ ms, see \cite{PK94,KU95}). The proposed
EOSs are qualitatively and quantitatively very different from each
other. Some are based on relativistic many-body theories while others
use nonrelativistic theories with baryon-baryon interaction
potentials. A classic collection of early proposed EOSs was compiled
by Arnett and Bowers \cite{AB}, while recent EOSs are described in
Salgado et al.  \cite{S94}.

High density equations of state with pion condensation have
been proposed by Migdal \cite{Mi71} and Sawyer and Scalapino
\cite{SS72}.
The possibility of Kaon condensation is discussed by Brown
and Bethe \cite{BB94} and questioned by Pandharipande et
al. \cite{P95}. Many authors have examined the possibility of stars
composed of strange quark matter and a recent review can be found in
\cite{WSWG97}.  

The realistic EOSs are supplied in the form of an
energy density vs. pressure table and intermediate values
are interpolated. This results in some loss of accuracy because the
usual interpolation methods do not preserve thermodynamical
consistency. Recently however, Swesty \cite{S96} devised a cubic
Hermite interpolation scheme that does preserve thermodynamical
consistency  and the scheme has been shown to indeed produce higher
accuracy neutron star models in Nozawa et al. \cite{N97}.

\begin{itemize}
  \item {\bf Going further.} A discussion of hybrid stars, that have a
  mixed-phase region of quark and hadronic matter, can be found in 
  \cite{WGP97}. A study of the relaxation effect in dissipative 
  relativistic fluid theories is presented in \cite{Relax}.
\end{itemize}

\subsection{Numerical Schemes}

     Out of the ten components of the field equations that describe
the geometry of a rotating relativistic star, only four are
independent and one has the freedom to choose which four components to
use. After choosing four field equations, there are different methods
one can use to solve them. First models were obtained by Wilson
\cite{W72} and Bonazzola and Schneider \cite{BS74}.
Here we will review the following methods: Hartle's slow rotation 
formalism, the Newton-Raphson linearization scheme due to Butterworth and Ipser
\cite{BI76}, a scheme using Green's functions by Komatsu et
al. \cite{KEH89a,KEH89b}, a minimal surface scheme due to Wu et
al. \cite{W91}, and two spectral methods by Bonazzola et
al. \cite{BGSM93,BGM98}. Below we give a description about each method and
its various implementations (codes).

\subsubsection{Hartle}

     To $O(\Omega^2)$ the structure of a star changes only by
quadrupole terms and the equilibrium equations become a set of
ordinary differential equations. Hartle's \cite{H67,HT68} method
computes rotating stars in this slow-rotation approximation and a
review of slowly rotating models has been compiled by Datta
\cite{D88}. Weber et al. \cite{WG91}, \cite{WGW91} also implement
Hartle's formalism to explore the rotational properties of four new
EOSs. 

Weber and Glendenning \cite{WG92} attempt to improve on Hartle's
formalism in order to obtain a more accurate estimate of the angular
velocity at the mass-shedding limit but their models show large
discrepancies compared to corresponding models computed with fully
rotating schemes \cite{S94}.  Thus, Hartle's formalism cannot be used
to compute models of rapidly rotating relativistic stars with sufficient
accuracy.

\subsubsection{Butterworth and Ipser (BI)}

     The BI-scheme \cite{BI76} solves the four field equations
following a Newton-Raphson like linearization and iteration procedure.
One starts with a nonrotating model and increases the angular velocity
in small steps, treating a new rotating model as a linear perturbation
of the previously computed rotating model. Each linearized field
equation is discretized and the resulting linear system is solved. The
four field equations and the hydrostationary equilibrium equation are
solved separately and iteratively until convergence is achieved. 

The space is truncated at a finite distance from the star and the boundary
conditions there are imposed by expanding the metric potentials in
powers of $1/r$. Angular derivatives are approximated by high-accuracy
formulae and models with density discontinuities are treated specially
at the surface. An equilibrium model is specified by fixing its rest
mass and angular velocity.

     The original BI code was used to construct uniform density models
and polytropic models \cite{BI76,B76}. Friedman et al. \cite{FIP86, FIP89}, 
extend the BI code to obtain a large number of rapidly
rotating models based on a variety of realistic EOSs.  Lattimer et al.
\cite{L90} used a code which was also based on the BI scheme to
construct rotating stars using recent ``exotic'' and schematic EOSs,
including pion or Kaon condensation and self-bound strange quark
matter.

\subsubsection{Komatsu, Eriguchi and Hachisu (KEH)}

     In the KEH scheme \cite{KEH89a,KEH89b}, the same set of field
equations as in BI is used, but the three elliptic-type field
equations are converted into integral equations using appropriate
Green's functions. The boundary conditions at large distance from the
star are thus incorporated into the integral equations, 
but the region of integration is
truncated at a finite distance from the star.  The fourth field
equation is an ordinary first-order differential equation. The field
equations and the equation of hydrostationary equilibrium are solved 
iteratively, fixing the maximum energy density and the ratio of the polar 
radius to the equatorial radius, until
convergence is achieved. In \cite{KEH89a,KEH89b} and \cite{EHN94} the
original KEH code is used to construct uniformly and differentially
rotating stars for both polytropic and realistic EOSs.

     Cook, Shapiro and Teukolsky (CST) improve on the KEH scheme by
introducing a new radial variable which maps the semi-infinite region
$[0,\infty)$ to the closed region $[0,1]$. In this way, the region of
integration is not truncated and the model converges to a higher
accuracy. Details of the code are presented in \cite{CST92} and polytropic
and realistic models are computed in \cite{CST94a} and \cite{CST94b}.

     Stergioulas and Friedman (SF) implement their own KEH code following
the CST scheme. They improve on the accuracy of the code by a special
treatment of the second order radial derivative that appears in the
source term of the first-order differential equation for one of the
metric functions.  This derivative was introducing a numerical error
of $1 \% -2 \%$ in the bulk properties of the most rapidly rotating
stars computed in the original implementation of the KEH scheme. The
SF code is presented in \cite{SF95} and in \cite{SPHD}. It is
available as a public domain code, named {\tt rns}, and can be  
downloaded from \cite{RNS}.

\subsubsection{Wu et al. (WMSHR)}

     The numerical scheme by Wu et al. \cite{W91} implements the
minimal surface formalism for rotating axisymmetric space-times
\cite{NH84,N85,N88}, in which Einstein's field equations are
equivalent to the minimal surface equations in an abstract Riemannian
potential space with a well-defined metric, whose coordinates are the
four metric functions of the usual stationary, axisymmetric metric. A
finite element technique is used and the system of equations is solved
by a Newton-Raphson method. Models based on realistic EOSs are
presented in \cite{NH92}. The WMSHR code has been used to visualize
rapidly rotating stars by embedding diagrams and 4D-ray-tracing
pictures (see \cite{H93} for a review).

\subsubsection{Bonazzola et al. (BGSM)}

     In the BGSM scheme \cite{BGSM93}, the field equations are derived
in the $3+1$ formulation. All four equations describing the
gravitational field are of elliptic type. This avoids the problem with
the second-order radial derivative in the source term of the ODE used
in BI and KEH.  The equations are solved using a spectral method,
i.e. all functions are expanded in terms of trigonometric functions in
both the angular and radial directions and a Fast Fourier Transform
(FFT) is used to obtain coefficients. Outside the star a redefined
radial variable is used, which maps infinity to a finite distance.

     In \cite{SAL94} the code is used to construct a large number of
models based on recent EOSs. The accuracy of the computed models is
estimated using two general relativistic Virial identities, valid for
general asymptotically flat space-times, that were discovered by
Gourgoulhon and Bonazzola \cite{GB94,BG94}. 

While the field equations used in the BI and KEH schemes
assume a perfect fluid, isotropic
stress-energy tensor, the BGSM formulation makes no assumption about
the isotropy of $T_{ab}$. Thus, the BGSM code can compute stars with
magnetic field, solid crust or solid interior and it can also be used
to construct rotating boson stars. 

Since it is based on the $3+1$
formalism, the BGSM code is also suitable for providing high-accuracy,
unstable equilibrium models as initial data for an axisymmetric
collapse computation.

\subsubsection{Bonazzola et al. (BGM-98)}

     The BGSM spectral method has been improved by Bonazzola et al.
\cite{BGM98} allowing for several domains of integration. One of the domain 
boundaries is chosen to coincide with the surface of the star and
a regularization procedure is introduced for the infinite 
derivatives at the surface (that appear in the density field when stiff 
equations of state are used). This allows models to be computed that are
free of Gibbs phenomena at the surface. The method is also 
suitable for constructing quasi-stationary models of binary neutron
stars.

\subsubsection{Direct Comparison of Numerical Codes}

     The accuracy of the above numerical codes can be estimated, if one
constructs exactly the same models with different codes and compares
them directly. The first such comparison of rapidly rotating models
constructed with the FIP and SF codes is presented by Stergioulas and
Friedman in \cite{SF95}.
Rapidly rotating models constructed with several EOS's agree to $0.1 \% -
1.2 \%$ in the masses and radii and to better than $2 \%$ in any other
quantity that was compared (angular velocity and momentum, central
values of metric functions etc.). This is a very satisfactory agreement,
considering that the BI code was using relatively few grid points, due
to limitations of computing power at the time of its implementation. 

In \cite{SF95}, it is also shown that a large
discrepancy between certain rapidly rotating models, constructed with
the FIP and KEH codes, that was reported by Eriguchi et
al. \cite{EHN94}, was only due to the fact that a different version of
a tabulated EOS was used in \cite{EHN94} than by FIP.

   Recently, Nozawa et al. \cite{N97} have completed an extensive
direct comparison of the BGSM, SF and the original KEH codes, using a
large number of models and equations of state. More than twenty
different quantities for each model are compared and the relative
differences range from $10^{-3}$ to $10^{-4}$ or better, for 
smooth equations of state. The agreement
is excellent for soft polytropes, which shows that all three codes are
correct and compute the desired models to an accuracy that
depends on the number of grid-points used to represent the
spacetime. 

If one makes the extreme assumption of uniform density, the agreement is
at the level of $10^{-2}$. In the BGSM code this is due to the fact
that the spectral expansion in terms of trigonometric functions cannot
accurately represent functions with discontinuous first-order
derivatives at the surface of the star. In the KEH and SF codes, the
three-point finite-difference formulae cannot accurately represent
derivatives across the discontinuous surface of the star. 

The accuracy of the three codes is also estimated by the use of the two Virial
identities due to Gourgoulhon and Bonazzola \cite{GB94,BG94}.
Overall, the BGSM and SF codes show a better and more consistent
agreement than the KEH code with BGSM or SF. This is largely due to
the fact that the KEH code does not integrate over the whole spacetime
but within a finite region around the star, which introduces some
error in the computed models.

\begin{itemize}
   \item {\bf Going further.} A review of spectral methods in general 
    relativity can be found in \cite{BFGM96}. A formulation for 
    nonaxisymmetric, uniformly rotating equilibrium configurations in the
    second post-Newtonian approximation is presented in \cite{AS96}.
\end{itemize}

\subsection{Properties of Equilibrium Models}

\subsubsection{Bulk Properties of Equilibrium Models}

     Neutron star models constructed with various realistic EOSs have
considerably different bulk properties, due to the large uncertainties
in the equation of state at high densities. Very compressible (soft)
EOSs produce models with small maximum mass, small radius, and large
rotation rate. On the other hand, less compressible (stiff) EOSs
produce models with a large maximum mass, large radius, and low
rotation rate. 

The gravitational mass,
equatorial radius and rotational period of the maximum mass model
constructed with one of the softest EOSs (EOS B) ($1.63M_{\odot}$,
9.3km, 0.4ms) are a factor of two smaller than the mass, radius and
period of the corresponding model constructed by one of the stiffest
EOSs (EOS L) ($3.27M_{\odot}$, 18.3km, 0.8ms). The two models differ
by a factor of 5 in central energy density and a factor of 8 in the
moment of inertia! 

Not all properties of the maximum mass models
between proposed EOSs differ considerably. For example, most realistic 
EOSs predict a maximum mass model with a ratio of rotational to
gravitational energy $T/W$ of $0.11 \pm 0.02$, a dimensionless angular
momentum $cJ/GM^2$ of $0.64 \pm 0.06$ and an eccentricity of $0.66 \pm
0.04$, \cite{FI92}. Hence, between the set of realistic EOSs, some 
properties are
directly related to the stiffness of the EOS while other properties
are rather insensitive to stiffness.

    Compared to nonrotating stars, the effect of rotation is to
increase the equatorial radius of the star and also to increase the
mass that can be sustained at a given central energy density. As a
result, the mass of the maximum mass rotating model is roughly $15 \% -20
\%$ higher than the mass of the maximum mass nonrotating model, for 
typical realistic EOSs.  The corresponding increase in radius is $30 \% 
-40 \%$.

The deformed shape of a rapidly rotating star creates a distortion, away
from spherical symmetry, in its gravitational field. Far from the star,
the distortion is measured by the quadrupole-moment tensor
$Q_{ab}$. For uniformly rotating, axisymmetric and equatorially symmetric
configurations, one can define a scalar quadrupole moment $Q$, which
can be extracted from the asymptotic expansion, at large r, of the metric
function $\nu$.

Laarakkers and Poisson \cite{LP97}, numerically compute the scalar
quadrupole moment $Q$ for several equations of state, using the rotating
neutron star code {\tt rns} \cite{RNS}. They find that for fixed 
gravitational mass $M$, the quadrupole moment is given as a simple 
quadratic fit
\begin{equation}
      Q = -a \frac{J^2}{ Mc^2},
\end{equation}
where $J$ is the angular momentum of the star and $a$ is a dimensionless
quantity that depends on the equation of state. The above quadratic fit 
reproduces $Q$ with a remarkable accuracy. The quantity $a$ varies between
$a \sim 2$ for very soft EOSs and $a \sim 8$ for very stiff EOSs, for
$M=1.4 M_{\odot}$ neutron stars.

     For a given zero-temperature EOS, the uniformly rotating
equilibrium models form a 2-dimensional surface in the 3-dimensional
space of central energy density, gravitational mass and angular momentum
\cite{SF95}.  
The surface is limited by the nonrotating models
($J=0$) and by the models rotating at the mass-shedding (Kepler) limit, 
i.e. at the maximum allowed angular velocity so that the star does 
not shed mass
at the equator.  
Cook et al. \cite{CST92,CST94a,CST94b} have shown that the model with 
maximum angular velocity
does not coincide with the maximum mass model, but is generally very
close to it in central density and mass. Stergioulas and Friedman
\cite{SF95} show that the maximum angular velocity and maximum baryon
mass equilibrium models are also distinct. The distinction becomes
significant in the case where the EOS has a large phase transition
near the central density of the maximum mass model, otherwise the
models of maximum mass, baryon mass, angular velocity and angular
momentum can be considered to coincide for most purposes.

\subsubsection{An Empirical Formula for the Kepler Velocity}

     In the Newtonian limit the maximum angular velocity of uniformly
rotating polytropic stars is, $\Omega_{max} \simeq (2/3)^{3/2} (GM/R^3)^{1/2}$
(see \cite{ST83}).  
For relativistic stars, the empirical
formula \cite{HZ89,FIP89,Fr89}
\begin{equation}
\Omega_{max} = 0.67 \sqrt{\frac{G M_{max}}{R_{max}^3}},
                 \label{e:empirical}
\end{equation}
gives the maximum angular velocity in terms of the mass and radius of
the maximum mass {\em nonrotating} model with an accuracy of $5 \% -7 \%$,
without actually having to construct rotating models. 

The empirical
formula results from universal proportionality relations that exist
between the mass and radius of the maximum mass rotating model and
those of the maximum mass nonrotating model for the same EOS. Lasota
et al. \cite{LHA96} find that, for most EOSs, the coefficient in the empirical
formula is an almost linear function of the parameter
\begin{equation}
\chi_s = \frac{2GM_{max}}{R_{max} c^2}, 
\end{equation}
When this relation is taken into account in the empirical formula, it
reproduces the exact values with a relative error of only $1.5 \%$. 

Weber
and Glendenning \cite{WG91,WG92}, try to reproduce analytically the
empirical formula in the slow rotation approximation but the formula
they obtain involves the mass and radius of the maximum mass rotating
configuration, which is different from what is involved in 
(\ref{e:empirical}).

\subsubsection{The Upper Limit on Mass and Rotation}

     The maximum mass and minimum period of rotating relativistic
stars computed with realistic EOSs from the Arnett and Bowers
collection \cite{AB} are about $3.3 M_{\odot}$ (EOS L) and $0.4$ms
(EOS B), while $1.4 M_{\odot}$ neutron stars, rotating at the Kepler
limit, have a rotational periods between $0.53$ms (EOS B) and 1.7ms
(EOS M) \cite{CST94b}. The maximum, accurately measured, neutron star 
mass is
currently $1.44 M_{\odot}$, but there are also indications for $2.0
M_{\odot}$ neutron stars \cite{KFC97}. The minimum observed pulsar
period is 1.56ms \cite{KU95}, which is close to the experimental 
sensitivity of
recent pulsar searches (an ongoing experiment is designed to detect
sub-millisecond pulsars, if they exist \cite{DA96}). 

In principle,
neutron stars with maximum mass or minimum period could exist, if they 
are born
as such in a core collapse, or if they accrete the right amount of
matter and angular momentum during an accretion-induced spin-up
phase. Such a phase could also follow the creation of an $1.4
M_{\odot}$ neutron star during the accretion induced collapse of a
white dwarf. 

In reality, only a very small fraction, if any, of neutron
stars will be close to the maximum mass or minimum period limit. In
addition, rapidly rotating nascent neutron stars are subject to 
a nonaxisymmetric instability, which
lowers their initial rotation rate and neutron stars
with a strong magnetic field have their rotation rate limited by
the Kepler velocity at their Alfven radius, where the accretion
pressure balances the magnetospheric pressure \cite{KU95}.

\begin{itemize}
  \item {\bf Going further.} A recent review by J. L. Friedman on the
  upper limit on rotation of relativistic stars can be found in 
  \cite{Fr95}.
\end{itemize}

\subsubsection{The Upper Limit on Mass and Rotation Set by Causality}

     Current proposed EOSs are reliable only to about twice nuclear
density and result in very different values for the maximum mass and
minimum period of neutron stars.  If one is interested in obtaining
upper limits on the mass and rotation rate, independent of the proposed
EOSs, one has to rely on fundamental physical principles. 

Instead of using realistic EOSs, one constructs a set of artificial
EOSs that satisfy only a minimal set of physical constraints, which
represent what we know about the equation of state of matter with high
confidence. One then searches among all these EOSs to obtain the one
that gives the maximum mass or minimum period. The minimal set of
constraints that have been used in such searches are that
\begin{enumerate}
   \item the high density EOS matches to the known low density EOS at
         some matching energy density $\epsilon_m$,
   \item the matter at high densities satisfies the causality
         constraint (the speed of sound is less than the speed of light).
\end{enumerate}
In relativistic perfect fluids, the speed of sound is the
characteristic velocity of the fluid evolution equations and the
causality constraint translates into the requirement
\begin{equation}
           dp/d \epsilon \leq 1.
\end{equation}
(see e.g. Geroch and Lindblom \cite{GL}). It is assumed that the 
fluid will still behave as a perfect fluid when it is perturbed 
from equilibrium.

     For nonrotating stars, Rhoades and Ruffini showed that the EOS
that satisfies the above two constraints and yields the maximum mass
consists of a high density region as stiff as possible (i.e. at the
causal limit, $dp/d \epsilon=1$), that matches directly to the known low
density EOS. For a chosen matching density $\epsilon_m$, they computed
a maximum mass of $3.2 M_{\odot}$. However, this is not the
theoretically maximum mass of nonrotating neutron stars, as is often
quoted in the literature. Hartle and Sabbadini \cite{HS77} point out
that $M_{max}$ is sensitive to the matching energy density and Hartle
\cite{H78} computes $M_{max}$ as a function of $\epsilon_m$.
\begin{equation}
     M_{max} = 4.8 \ \Bigl( \frac{2 \times 10^{14} {\rm gr/cm}^3} 
                   { \epsilon_m} \Bigr)^{1/2} M_{\odot}.
\end{equation}

     In the case of rotating stars, Friedman and Ipser \cite{FI87}
assume that the absolute maximum mass is obtained by the same EOS as
in the nonrotating case and compute $M_{max}$ as a function of matching
density, assuming the BPS EOS holds at low densities. Stergioulas and
Friedman \cite{SF95} recompute $M^{rot}_{max}$ for rotating stars using 
the more recent FPS EOS at low densities, obtaining very nearly the same
result
\begin{equation}
     M^{rot}_{max} = 6.1 \ \Bigl( \frac{2 \times 10^{14} {\rm gr/cm}^3} 
                   { \epsilon_m} \Bigr)^{1/2} M_{\odot},
\end{equation}
where, $2 \times 10^{14} {\rm gr/cm}^3$ is roughly nuclear saturation 
density for the FPS EOS.

     A first estimate of the absolute minimum period of uniformly
rotating, gravitationally bound stars was computed by Glendenning
\cite{G92} by constructing nonrotating models and using the empirical
formula (\ref{e:empirical}) to estimate the minimum period. 
Koranda, Stergioulas and
Friedman \cite{KSF97} improve on Glendenning's results by constructing
accurate rapidly rotating models and show that Glendennings results
are accurate to within the accuracy of the empirical formula. 

Furthermore, they show that the EOS satisfying the minimal set
of constraints and yielding the minimum period star consists of a high
density region at the causal limit, which is matched to the known low
density EOS through an intermediate constant pressure region (that
would correspond to a first-order phase transition). Thus, the
EOS yielding absolute minimum period models is as stiff as possible at 
the central density
of the star (to sustain a large enough mass) and
as soft as possible in the crust, in order to have the smallest
possible radius (and rotational period). 

The absolute minimum period
of uniformly rotating stars is an (almost linear) function of the
maximum observed mass of nonrotating neutron stars
\begin{equation}
     P_{min} = 0.28 {\rm ms} + 0.2 (M_{max}^{nonrot.} - 1.44 M_{\odot}),
\end{equation}
and is rather insensitive to the matching density $\epsilon_m$ (the
above result was computed for a matching number density
of $0.1 {\rm fm}^{-3}$).

     In \cite{KSF97}, it is also shown that an absolute limit on the
minimum period exists even without requiring that the EOS matches to a
known low density EOS (this is not true for the limit on the maximum
mass). Thus, using causality as the only constraint on the EOS, 
$P_{min}$ is lowered by only $3 \%$, which shows that the currently
known part of the nuclear EOS plays a negligible role in determining
the absolute upper limit on the rotation of uniformly rotating,
gravitationally bound stars.

\subsubsection{Spin-Up Prior to Collapse}

     Since rotation increases the mass that a neutron star of given
central density can support, there exist sequences of neutron stars
with constant baryon number that have no nonrotating member. Such
sequences are called supra-massive as opposed to normal sequences that
do have a nonrotating member. A nonrotating star can become
supra-massive by accreting matter and spinning-up to large rotation
rates; in another scenario, neutron stars could be born
supramassive after a core collapse. A supra-massive star evolves along
a sequence of constant baryon mass, slowly loosing angular
momentum. Eventually, the star reaches a point where it becomes
unstable to axisymmetric perturbations and collapses to a black hole.
The instability grows on a secular timescale, in the sense that it is
limited by the time required for viscosity to redistribute the star's
angular momentum. This timescale is comparable with the spin-up time
following a glitch \cite{FI92}.

Cook et al.
\cite{CST92,CST94a,CST94b} have discovered that a supramassive star 
approaching the axisymmetric instability, will actually
spin-up before collapse, even though it looses angular momentum.
This, potentially observable, effect is independent of the equations
of state and it is more pronounced for rapidly rotating massive
stars. In a similar phenomenon, normal stars can spin-up by loss of
angular momentum near the Kepler limit, if the equation of state is
extremely stiff or extremely soft.

\subsubsection{Rotating Magnetized Neutron Stars}

     The presence of a magnetic field was ignored in the
models of rapidly rotating relativistic stars that were considered in
the previous sections. The reason is that the observed surface dipole
magnetic field strength of pulsars ranges between $B=10^8$ G and $B=2
\times 10^{13}$ G. These values of $B$ imply a magnetic
field energy density that is too small compared to the energy density
of the fluid, to significantly affect the structure of a neutron
star. However, one cannot exclude the existence of neutron stars with
higher magnetic field strengths or the possibility that neutron stars
are born with much stronger magnetic fields, which then decay to the
observed values (Of course there are also many arguments against
magnetic field decay in neutron stars \cite{PK94}).  In addition, even
though moderate magnetic field strengths do not alter the bulk
properties of neutron stars, they may have an effect on the
damping or growth rate of various perturbations of an equilibrium
star, affecting its stability. For these reasons, a fully relativistic
description of magnetized neutron stars is desirable and, in fact,
Bocquet et al. \cite{BBGN95} achieved the first numerical computation
of such configurations. Here we give a brief summary of their work:

     A magnetized relativistic star in equilibrium can be described by
the coupled Einstein-Maxwell field equations for stationary,
axisymmetric rotating objects with internal electric currents. The
stress-energy tensor includes the electromagnetic energy density and
is non-isotropic (in contrast to the isotropic perfect fluid stress-
energy tensor). The equilibrium of the matter is given not only by the
balance between the gravitational force and the pressure gradient, but
the Lorentz force due to the electric currents also enters the
balance. For simplicity, Bocquet et al.  consider only poloidal
magnetic fields, which preserve the circularity of the space-time.
Also, they only consider stationary configurations, which excludes
magnetic dipole moments non-aligned with the rotation axis, since in
that case the star emits electromagnetic and gravitational waves.  The
assumption of stationarity implies that the fluid is necessarily
rigidly rotating (if the matter has infinite conductivity) \cite{BGSM93}. 
Under these assumptions, the
electromagnetic field tensor $F^{ab}$ is derived from a potential
1-form $A_a$ with only two non-vanishing components, $A_t$ and
$A_{\phi}$, which are given by a scalar Poisson and a vector
Poisson equation respectively. Thus, the two equations describing the
electromagnetic field are of similar type as the four field equations
that describe the gravitational field.

 The construction of magnetized models with $B<10^{13}$ G
confirms that magnetic fields of this strength have a negligible
effect on the structure of the star. However, if one increases the
strength of the magnetic field above $10^{14}$ G, one observes
significant effects, such as a flattening of the star. The magnetic
field cannot be increased indefinitely, but there exists a maximum
value of the magnetic field strength, of the order of $10^{17}$ G, for
which the magnetic field pressure at the center of the star equals the
fluid pressure. Above this value, the fluid pressure decreases more
rapidly away from the center along the symmetry axis, than the magnetic
pressure. Instead of pressure, there is tension along the symmetry
axis and no stationary configuration can exist.

The shape of a strongly magnetized star is flattened because the 
 Lorentz forces exerted by the E/M field on the fluid
act as centrifugal forces. A star with a magnetic field near the
maximum value for stationary configurations, displays a
pinch along the symmetry axis, because there, the magnetic pressure
exceeds the fluid pressure. The maximum fluid density inside the star
is not attained at the center, but away from it. The presence of a
strong magnetic field also allows a maximum mass configuration with
larger $M_{max}$ than for the same EOS with no magnetic field and this
is in analogy with the increase of $M_{max}$ induced by rotation. For
nonrotating stars, the increase in $M_{max}$, due to a strong magnetic
field, is $13 \% - 29 \%$, depending on the EOS.
Following the increase in mass, the maximum allowed angular
velocity for a given EOS also increases in the presence of a magnetic
field.

     Bocquet et al. are planning to use their code in the study of two 
types of possible instabilities in magnetized neutron stars, i) a pure 
E/M instability towards another electric current/magnetic field
distribution of lower energy and ii) a nonaxisymmetric instability for
rapidly rotating models, which would be the analog of a Jacobi-type
transition in non-magnetized stars. In perfect fluid models with a magnetic
field, one would also expect a CFS-instability driven by electromagnetic
waves.

\subsubsection{Rapidly Rotating Proto-Neutron Stars}

Following the gravitational collapse of a massive stellar core, a
proto-neutron star (PNS) is born. Initially it has a large radius of
about 100km and a temperature of 50-100MeV. The PNS may be born with a
large rotational kinetic energy and initially it will be
differentially rotating. Due to the violent nature of the
gravitational collapse, the PNS pulsates heavily, emitting significant
amounts of gravitational radiation. After a few hundred pulsational
periods, bulk viscosity will damp the pulsations significantly.
Rapid cooling due to deleptonization transforms the PNS to a hot
neutron star of $T \sim 10$MeV shortly after its formation. In addition,
viscosity reduces the differential rotation to a nearly uniform rotation on a
timescale of seconds \cite{GHZ98} and the neutron star becomes quasi-stationary.
Since the details of the PNS evolution determine the exact properties
of the resulting cold NSs, proto-neutron stars must be modeled
realistically in order to understand the structure of cold neutron
stars.

     Hashimoto et al. \cite{HOE94} and Goussard et al. \cite{GHZ96}
recently constructed fully relativistic models of rapidly rotating,
hot proto-neutron stars. The authors use finite-temperature EOSs
\cite{O93,LS91}, to model the
interior of PNSs. Important parameters, which determine
the local state of matter but are largely unknown, are the lepton
fraction $Y_l$ and the temperature profile. Hashimoto et
al. consider only the limiting case of zero lepton
fraction $Y_l=0$ and classical isothermality, while Goussard et
al. consider several non-zero values for $Y_l$ and two different
limiting temperature profiles - a constant entropy profile and a
relativistic isothermal profile. In both \cite{HOE94} and \cite{O93},
differential rotation is neglected to a first approximation.

     The construction of numerical models with the above assumptions
shows that, due to the high temperature and the presence of trapped
neutrinos, PNSs have a significantly larger radius than cold NSs.
These two effects give the PNS an extended envelope which, however,
contains only roughly $0.1 \%$ of the total mass of the star. This
outer layer cools more rapidly than the interior and becomes
transparent to neutrinos, while the core of the star remains hot and
neutrino opaque for a longer time. The two regions are separated by
the ``neutrino sphere''.

     Compared to the $T=0$ case, an isothermal EOS with temperature of
25MeV has a maximum mass model of only slightly larger mass. In
contrast, an isentropic EOS with a nonzero trapped lepton number
features a maximum mass model that has a considerably lower mass than
the corresponding model in the $T=0$ case and a stable PNS transforms
to a stable neutron star. If, however, one considers the hypothetical
case of a large amplitude phase transition which softens the cold EOS
(such as a Kaon condensate), then $M_{max}$ of cold neutron stars is
lower than $M_{max}$ of PNSs and a stable PNS with maximum mass will
collapse to a black hole after the initial cooling period. This
scenario of delayed collapse of nascent neutron stars has been
proposed by Brown and Bethe \cite{BB94} and investigated by Baumgarte
et al. \cite{BST96}.

     An analysis of radial stability of PNSs \cite{GHZ97} shows that,
for hot PNSs, the maximum angular velocity star almost coincides with
the maximum mass star, as is also the case for cold EOSs.

     Because of their increased radius, PNSs have a different 
mass-shedding limit than cold NSs. For an isothermal profile, 
the mass-shedding limit proves to be
sensitive to the exact location of the neutrino sphere. For the EOSs
considered in \cite{HOE94} and \cite{GHZ96} PNSs have a maximum
angular velocity that is considerably less than the maximum angular
velocity allowed by the cold EOS. Stars that have nonrotating
counterparts (i.e. that belong to a normal sequence) contract and
speed up while they cool down. The final star with maximum rotation is
thus closer to the mass-shedding limit of cold stars than was the hot
PNS with maximum rotation.  Surprisingly, stars belonging to a
supra-massive sequence exhibit the opposite behavior. If one assumes
that a PNS evolves without loosing angular momentum or accreting mass,
then a cold neutron star produced by the cooling of a hot PNS has a
smaller angular velocity than its progenitor. This purely relativistic
effect was pointed out in \cite{HOE94} and confirmed in
\cite{GHZ96}. It should be noted here, that a small amount of differential
rotation significantly affects the mass-shedding limit, allowing more
massive stars to exist than uniform rotation allows. Taking differential
rotation into account, a more recent study by Goussard et al. \cite{GHZ98} 
suggests that proto-neutron stars created in a gravitational collapse
cannot spin faster than 1.7 ms.

\section{Oscillations and Stability}

     The study of oscillations of relativistic stars has the potential
of yielding important information about both the bulk properties and
the composition of the interior of the star i.e.  about the equation
of state of matter at very high densities, in about the same way that
helioseismology is providing us with information about the interior of
the Sun.  In a neutron star - accretion disk system, the star-disk
interaction can drive oscillations and one of the possible explanations
for kHz quasi-periodic oscillations recently discovered in several X-ray sources 
are neutron star pulsations \cite{vdK97} (for an early proposal that
such oscillations may be observable, see \cite{MVHB85}).

     Neutron star pulsations may be a detectable source of
gravitational radiation. The pulsations can be excited after a core
collapse or during the final stages of a neutron star binary system 
coalescence. Rapidly rotating neutron stars are unstable to the emission of 
detectable gravitational
waves for a short time after their formation. The identification of
gravitational waves produced by a neutron star can lead to the
determination of its mass and radius and several such
determinations can help reconstruct the equation of state of matter at
very high energy densities \cite{Ko97}.

     The oscillations of relativistic stars are actually a non-linear
phenomenon and their numerical computation would require a full 3-D
relativistic hydrodynamics code, which is not yet available.  However,
apart from the initial oscillations following core collapse, the
oscillations of an equilibrium star are of small magnitude compared to
its radius and it will suffice to approximate them as linear
perturbations. Such perturbations can be described in two equivalent
ways.  In the Lagrangian approach, one studies the changes in a given
fluid element as it oscillates about its equilibrium position.  In the
Eulerian approach, one studies the change in fluid variables at a
fixed point in space.  Both approaches have their strengths and
weaknesses. 

In the Newtonian limit, the Lagrangian approach has been
used to develop variational principles \cite{LBO67,FS78} but the
Eulerian approach proved to be more suitable for numerical
computations of mode frequencies and eigenfunctions 
 \cite{IM85,M85,IL90,IL91a,IL91b}. Clement \cite{C81} 
used the Lagrangian approach to
obtain axisymmetric normal modes of rotating stars, while nonaxisymmetric
solutions were obtained in the Lagrangian approach by Imamura et al.
\cite{IFD85} and in the Eulerian approach by Managan \cite{M85} and
Ipser and Lindblom \cite{IL90}.

\subsection{Quasi-Normal Modes of Oscillation}

     The spacetime of a nonrotating star is static and spherically
symmetric. A general linear perturbation can be written as a sum of
quasi-normal modes that are characterized by the indices $(l,m)$ of the
spherical harmonic $Y_l^m$ and have angular and time-dependence of the
form
\begin{equation}
   \delta Q  \sim f(r) Y_l^m(cos\theta ) e^{i\omega_p t},
\end{equation}
where $Q$ is a scalar unperturbed quantity, $\omega_p$ is the angular 
frequency
of the mode, as measured by a distant inertial observer and $f(r)$
represents the radial dependence of the perturbation. Normal modes of
nonrotating stars are degenerate in m and it suffices to study the
axisymmetric $(m=0)$ case. 

The perturbation of the metric, $\delta g_{ab}$,
can be expressed in terms of spherical, vector and tensor harmonics.
These are either of "polar" or "axial" parity.  Here, parity is
defined as the change in sign under a combination of reflection in
the equatorial plane and rotation by $\pi$. A polar perturbation has
parity $(-1)^l$, while an axial perturbation has parity $(-1)^{l+1}$.
Because of the spherical background, the polar and axial
perturbations of a nonrotating star are completely decoupled. 

A normal mode solution satisfies the perturbed gravitational field
equations
\begin{equation}
            \delta(G^{ab}-8 \pi T^{ab})=0,
\end{equation}
and the perturbation of the conservation of the stress-energy tensor
\begin{equation}
            \delta(\nabla_aT^{ab})=0.
\end{equation}
For given $(l,m)$, a solution exists for any value of the
eigenfrequency $\omega_p$ and it consists of ingoing- and outgoing-wave
parts. Outgoing modes are defined by the discrete set of
eigenfrequencies for which there are no incoming waves at infinity.
These are the modes that will be excited in various astrophysical
situations.

     The main modes of pulsation that are known to exist in 
relativistic stars have been classified as follows ($f_0$ and $\tau_0$ are
typical frequencies and damping times of the most important modes 
in the nonrotating limit):
\begin{enumerate}
  \item {\it Polar fluid modes}

     Are slowly damped modes analogous to the Newtonian fluid
      pulsations:
   \begin{itemize}  
    \item  $f$(undamental)-mode: surface mode due to the interface between
        the star and its surroundings ($f_0 \sim 2$khz, $\tau_0<1$sec),

    \item  $p$(ressure)-modes: nearly radial ($f_0 >4$kHz, 
           $\tau_0 > 1$s),

    \item  $g$(ravity modes): nearly tangential, only exist for finite
       temperature stars ($f_0<500$Hz, $\tau_0 > 5$s).
   \end{itemize}
  \item {\it Axial fluid modes}

  \begin{itemize}  
    \item   $r$(otation) modes: degenerate at zero-frequency for
       nonrotating stars. In a rotating star, generically
       unstable. Frequencies from zero to kHz,
       growth times inversely proportional to a high power of the
       star's angular velocity.
   \end{itemize}  

  \item {\it Polar and axial spacetime modes}

   \begin{itemize}  
    \item $w$(ave)-modes: Analogous to the quasi-normal modes of a black hole.
       High frequency, strongly damped modes ($f_0>6$kHz, 
       $\tau_0 \sim 0.1$msec).
   \end{itemize}  
\end{enumerate}
For a more detailed description of various modes see
\cite{KB97,KLH96,MacD88,CAR86,KScha98}.

\subsection{Effect of Rotation on Quasi-Normal Modes}

     In a continuous sequence of rotating stars, a quasi-normal mode of
index $l$ is defined as the mode which, in the nonrotating limit,
reduces to the quasi-normal mode of the same index $l$. Rotation has several
effects on the modes of a previously nonrotating star:
\begin{enumerate}
 \item The degeneracy in the index $m$ is removed and a nonrotating mode
      of index $l$ is split into $2l+1$ different $(l,m)$ modes.

 \item Prograde ($m<0$) modes are now different than retrograde ($m>0$)
      modes.

 \item A rotating ``polar" $l$-mode consists of a sum of purely polar
      and purely axial terms \cite{SPHD}
\begin{equation}
         P_l^{rot} \sim \sum_{l'=0}^\infty(P_{l+2l'} +A_{l+2l' \pm 1}),
\end{equation}
      that is, rotation couples a polar
      $l$-term to an axial $l \pm 1$ term (the coupling to the $l+1$ term
      is, however, strongly favored over the coupling to the $l-1$ term 
      \cite{CF91}).
      Similarly, for a rotating ``axial'' mode,
\begin{equation}
         A_l^{rot} \sim \sum_{l'=0}^\infty(A_{l+2l'} +P_{l+2l' \pm 1}),
\end{equation}
 \item Frequencies and damping times are shifted. In general,
      frequencies (in the inertial frame) of prograde modes increase,
      while those of retrograde modes decrease with increasing rate
      of rotation.
\end{enumerate} 

     In rotating stars, quasi-normal modes of oscillation have only been
studied in the slow-rotation limit, in the post-Newtonian and in the
Cowling Approximations. The solution of the fully-relativistic
perturbation equations for a rapidly rotating star is still a very
challenging task and only recently they have been solved for zero-
frequency (neutral) modes \cite{SPHD,SF97}.

\begin{itemize}
   \item {\bf Going further.} The equations that describe oscillations
    of the solid crust of a rapidly rotating relativistic star are derived
   by Priou in \cite{Pr92}. The effects of superfluid hydrodynamics on the
   oscillations of neutron stars are investigated by Lindblom and Mendell 
   in \cite{LM94}.
\end{itemize}

\subsection{Axisymmetric Perturbation}

     Along a sequence of nonrotating relativistic stars with
increasing central energy density, there is always a model for which
the mass becomes maximum. The maximum mass turning point marks the
onset of a secular instability in the fundamental axisymmetric
pulsation mode of the star. 

Applying the turning point theorem
provided by Sorkin \cite{So82}, Friedman Ipser and Sorkin \cite{FIS88}
show that in the case of rotating stars the secular axisymmetric
instability sets in when the mass becomes maximum along a sequence of
constant angular momentum. An equivalent criterion is provided by Cook
et al. \cite{CST92} : the secular axisymmetric instability sets in when
the angular momentum becomes minimum along a sequence of constant rest
mass. 

The instability develops on a timescale that is limited by the
time required for viscosity to redistribute the star's angular
momentum. This timescale is long compared to the dynamical timescale
and comparable to the spin-up time following a pulsar glitch. When it
becomes secularly unstable, a star evolves in a quasi-stationary
fashion until it encounters the dynamical instability and collapses to
a black hole. Thus, the onset of the secular instability to axisymmetric
perturbations separates stable neutron stars from neutron stars that
will collapse to a black hole.

     Goussard et al. \cite{GHZ96} extend the stability criterion to hot
protoneutron stars with nonzero total entropy. In this case, the loss
of stability is marked by the configuration with minimum angular
momentum along a sequence of both constant rest mass and total
entropy. 

In the nonrotating limit, Gondek et al. \cite{GHZ97} compute
frequencies and eigenfunctions of axisymmetric pulsations of hot
proto-neutron stars and verify that the secular instability sets in at
the maximum mass turning point, as is the case for cold neutron stars.

\begin{itemize} 
   \item {\bf Going further} The stabilization of a relativistic star, that
   is marginally stable to axisymmetric perturbations, by an external
   gravitational field, is discussed in \cite{Th97}.
\end{itemize}  

\subsection{Nonaxisymmetric Perturbations}

\subsubsection{Nonrotating Limit}

     For a spherical star, it suffices to study the $m=0$ axisymmetric
modes of pulsation, since the $m \neq 0$ modes can be obtained by a
rotation of the coordinate system.

Thorne, Campolattaro and Price, in a series of papers
\cite{TC67,PT69,Th69}, initiated the computation of nonradial modes by
formulating the problem in the Regge-Wheeler (RW) gauge \cite{RW} and
numerically computing nonradial modes for a number of neutron star
models. A variational method for obtaining eigenfrequencies and
eigenfunctions has been constructed by Detweiler and Ipser
\cite{DI73}. Lindblom and Detweiler \cite{LD83} explicitly reduced the
system of equations to four first-order ordinary differential
equations and obtained more accurate eigenfrequencies and damping
times for a larger set of neutron star models. They later realized
that their system of equations is sometimes singular inside the star
and obtained an improved set of equations which is free of this
singularity \cite{DL85}.

     Chandrasekhar and Ferrari \cite{CF91} express the
nonradial pulsations in terms of a fifth-order system in a diagonal
gauge, which is independent of fluid variables. They thus reformulate
the problem in a way analogous to the scattering of gravitational
waves off a black hole. Ipser and Price \cite{IP91} show that in the
RW gauge, nonradial pulsations can be described by a system of two
second-order equations, which can also be independent of fluid
variables. In addition, they find that the diagonal gauge of
Chandrasekhar and Ferrari has a remaining gauge freedom which, when
removed, also leads to a fourth-order system of equations \cite{PI91}.

     In order to locate purely outgoing-wave modes, one has to
be able to distinguish the outgoing-wave part from the ingoing-wave
part at infinity. In the Thorne et al. and Lindblom and Detweiler
schemes, this is achieved using analytic approximations of the
solution at infinity. 

$W$-modes pose a more challenging numerical
problem because they are strongly damped and the techniques used for $f$ and
$p$ modes fail to distinguish the outgoing-wave part, but
Andersson, Kokkotas and Schutz \cite{AKS95}, successfully combine a
redefinition of variables with a complex-coordinate integration method,
obtaining highly accurate complex
frequencies for $w$ modes. In this method, the ingoing and outgoing 
solutions are
separated by numerically calculating their analytic continuations to a
place in the complex-coordinate place, where they have comparable
amplitudes. Since this approach is purely numerical, it could prove
to be suitable for the computation of quasi-normal modes in rotating stars,
where analytic solutions at infinity are not available.

     The non-availability of asymptotic solutions at infinity in the
case of rotating stars is one of the major difficulties for computing
outgoing modes in rapidly rotating relativistic stars. A new
development that may help to overcome this problem, at least to an
acceptable approximation, is presented in \cite{LMI97} by Lindblom, Mendell
and Ipser.

     The authors obtain approximate near-zone boundary conditions for
the outgoing modes that replace the outgoing-wave condition at
infinity and that enable one to compute the eigenfrequencies with very
satisfactory accuracy. First, the
pulsation equations of polar modes in the Regge-Wheeler gauge are 
reformulated as a set
of two second-order radial equations for two potentials - one
corresponding to fluid perturbations and the other to the
perturbations of the spacetime. The
equation for the space-time perturbation reduces to a scalar wave
equation at infinity and to Laplace's equation for zero-frequency
solutions. From these, an approximate
boundary condition for outgoing modes is constructed and imposed in 
the near zone
of the star (in fact on its surface) instead at infinity. 
For polytropic models, the near-zone boundary condition yields
$f$-mode eigenfrequencies with real parts accurate to $0.01 \%-0.1 \%$
and imaginary parts with accuracy at the $10 \% -20 \%$ level, for the
most relativistic stars. If the near zone boundary condition can be
applied to the oscillations of rapidly rotating stars, the resulting
frequencies and damping times should have comparable accuracy.

\subsubsection{Slow Rotation Approximation}

     The slow rotation approximation has proven to be useful for
obtaining a first estimate of the effect of rotation on the pulsations
of relativistic stars. 
To lowest order in rotation, a polar 
$l$-mode of an
initially nonrotating star couples to an axial $l \pm 1$ mode in the
presence of rotation.  Conversely, an axial $l$-mode couples to a
polar $l \pm 1$ mode \cite{CF91}.

     The equations of nonaxisymmetric perturbations in the
slow-rotation limit and in the Regge-Wheeler gauge are derived by Kojima
in \cite{Koj92,Koj93}, where the complex frequencies $\sigma =
\sigma_R + i \sigma_I$ for the $l=m$ modes of various polytropes are
computed. For counterrotating modes, both $\sigma_R$ and $\sigma_I$
decrease, tending to zero, as the rotation rate increases (when
$\sigma$ passes through zero, the star becomes unstable to the
CFS-instability).  Extrapolating $\sigma_R$ and $\sigma_I$ to higher
rotation rates, Kojima finds a large discrepancy between the points
where $\sigma_R$ and $\sigma_I$ go through zero. This shows that the
slow rotation formalism cannot accurately determine the onset of the
CFS-instability of polar modes in rapidly rotating neutron stars. 

In \cite{Koj93b}, it is shown that, for slowly rotating stars,
 the coupling between polar and
axial modes affects the frequency of pulsation only to second order in
rotation, so that, in the slow rotation approximation, to $O( \Omega)$,
the coupling can be neglected when computing frequencies.

     The slow rotation approximation has also been used recently in
the study of the $r$-mode instability \cite{A97,Koj97}.

\subsubsection{Post-Newtonian Approximation}

     A first step towards the solution of the perturbation equations
in full relativity has been taken by Cutler and Lindblom 
\cite{Cu91,CL92,Li95} who obtain frequencies for the $l=m$ $f$-modes in
rotating stars in the first post-Newtonian (1-PN) approximation. The
perturbation equations are derived in the post-Newtonian formalism of
Gunnarsen \cite{G91}, i.e. the equations are separated into equations
of consistent order in $1/c$. 

Cutler and Lindblom show that in this
scheme, the perturbation of the 1-PN correction of the four-velocity
of the fluid can be obtained analytically in terms of other variables,
similarly to what is done for the perturbation in the four-velocity in
the Newtonian Ipser-Managan scheme.  
The perturbation in the 1-PN
corrections are obtained by solving an eigenvalue problem, which
consists of three second order equations, with the 1-PN correction
to the eigenfrequency of a mode, $\Delta \omega$, as the eigenvalue.

Cutler and Lindblom obtain a formula that yields $\Delta \omega$ if
one knows the 1-PN stationary solution and the solution to the
Newtonian perturbation equations. Thus, the frequency of a mode in the
1-PN approximation can be obtained without actually solving the 1-PN
perturbation equations numerically.  The 1-PN code was checked in the
nonrotating limit and it was found to reproduce the exact general
relativistic frequencies for stars with $M/R=0.2$ obeying an $N=1$
polytropic EOS with an accuracy of $3 \% - 8 \%$.

     Along a sequence of rotating stars, the frequency of a mode is
commonly described by the ratio of the frequency of the mode in the
comoving frame to the frequency of the mode in the nonrotating limit.
For an $N=1$ polytrope and for $M/R=0.2$, this frequency ratio is 
reduced by as much as $12 \%$ in the 1-PN approximation compared to 
its Newtonian counterpart (for the fundamental $l=m$ modes) which is
representative of the effect that general relativity has on the
frequency of quasi-normal modes in rotating stars.

\subsubsection{Cowling Approximation}

     In several situations, the frequency of pulsations in relativistic
stars can be estimated even if one completely neglects the
perturbation in the gravitational field, i.e. if one sets $\delta
g_{ab}=0$ in the perturbation equations \cite{MVS83}.  In this
approximation, the pulsations are described only by the perturbation
in the fluid variables and the scheme works quite well for $f$, $p$
and $r$-modes \cite{LS90}. A different version of the Cowling
approximation, in which $\delta g_{tr}$ is kept nonzero in
the perturbation equations, works better for $g$-modes \cite{Fi88}.

     Yoshida and Kojima \cite{YK97}, examine the accuracy of the
relativistic Cowling approximation in slowly rotating stars.  The
first-order correction to the frequency of a mode depends only
on the eigenfrequency and eigenfunctions of the mode in the absence of
rotation and on the angular velocity of the star. The eigenfrequencies
of $f$, $p_1$ and $p_2$ modes for slowly rotating stars with $M/R$
between 0.05 and 0.2 are computed (assuming polytropic EOSs with $N=1$
and $N=1.5$ and compared to their counterparts in the slow- rotation
approximation. 

For the $l=2$ $f$-mode, the relative error in the
eigenfrequency because of the Cowling approximation is $30 \%$ for
less relativistic stars ($M/R=0.05$) and about $15 \%$ for stars with
$M/R=0.2$ and the error decreases for higher $l$-modes. For the $p_1$
and $p_2$ modes the relative error is similar in magnitude but it is
smaller for less relativistic stars.  Also, for $p$-modes, the Cowling
approximation becomes more accurate for increasing radial mode
number. 

As an application, Yoshida and Eriguchi \cite{YE97} use
the Cowling approximation to estimate the onset of the CFS instability
in rapidly rotating relativistic stars.

\subsection{Nonaxisymmetric Instabilities}

\subsubsection{Introduction}

     Rotating cold neutron stars, detected as pulsars, have a
remarkably stable rotation period. But, at birth, or during accretion,
rapidly rotating neutron stars can be subject to various
nonaxisymmetric instabilities, which will affect the evolution of
their rotation rate

     If a protoneutron star has a sufficiently high rotation rate
(larger than $T/W \sim 0.27$ for uniformly rotating, constant density
Maclaurin spheroids), it will be subject to a dynamical instability
driven by hydrodynamics and gravity. Through the $l=2$ mode, the
instability will deform the star into a bar shape. This highly
nonaxisymmetric configuration will emit strong gravitational waves
with frequencies in the kHz regime. The development of the instability
and the resulting waveform have been computed numerically in the
context of Newtonian gravity and hydrodynamics by Houser et al.
\cite{HCS94}.

     At lower rotation rates, the star can become unstable to secular
nonaxisymmetric instabilities, driven by gravitational radiation or
viscosity. Gravitational radiation drives a nonaxisymmetric
instability when a mode that is retrograde with respect to the star
appears as prograde to a distant observer, via the
Chandrasekhar-Friedman-Schutz (CFS) mechanism \cite{C70,FS78}: A mode
that is retrograde in the corotating frame has negative angular
momentum, because the perturbed star has less angular momentum than
the unperturbed one. If, to a distant observer, the mode appears
prograde, it removes positive angular momentum from the star and thus
the angular momentum of the mode becomes increasingly negative. 

The instability evolves on a secular timescale, during which the star
loses angular momentum via the emitted gravitational waves. When the
star rotates slow enough, the mode becomes stable and the instability
proceeds on the longer timescale of the next unstable mode, unless it
is suppressed by viscosity.

     Neglecting viscosity, the CFS-instability is generic in rotating
stars for both polar and axial modes.  For polar modes, the instability
occurs only above some critical angular velocity, where the frequency
of the mode goes through zero in the inertial frame. 
The critical angular velocity is smaller for increasing mode
number $l$. Thus, there will always be a high enough mode number $l$,
for which a slowly rotating star will be unstable.  Axial modes 
are generically unstable in all rotating stars, since the mode
has zero frequency in the inertial frame when the star is nonrotating
  \cite{A97,FM97}. 

     The shear and bulk viscosity of neutron star matter is able to suppress
the growth of the CFS-instability except when the star passes through
a certain temperature window. In Newtonian gravity, it appears that the polar
mode CFS-instability can occur only in nascent neutron stars that
rotate close to the mass-shedding limit \cite{IL91a,IL91b,IL92,YE95,LM95}, 
but the determination of neutral $f$-modes in full
relativity \cite{SPHD,SF97} shows that relativity enhances the
instability, allowing it to occur in stars with smaller rotation rates 
than previously thought. 

\begin{itemize}
 \item {\bf Going further.} A new numerical method for the analysis of the
   ergoregion instability in relativistic stars, which may also be used
   for the analysis of nonaxisymmetric instabilities, is presented by
   Yoshida abd Eriguchi in \cite{YE96}.
\end{itemize}

\subsubsection{CFS-Instability of Polar Modes}

     The existence of the CFS-instability in rotating stars was first
demonstrated by Chandrasekhar \cite{C70} in the case of the $l=2$ mode
in uniformly rotating, constant density Maclaurin spheroids. Friedman
and Schutz \cite{FS78}, show that this instability also appears in
compressible stars and that all rotating self-gravitating perfect
fluid configurations are generically unstable to the emission of
gravitational waves. In addition, they find that a nonaxisymmetric
mode becomes unstable when its frequency vanishes in the inertial
frame. Thus, zero-frequency outgoing-modes in rotating stars are
neutral (marginally stable).

     In the Newtonian limit, neutral modes have been determined for
several polytropic EOSs \cite{IFD85,M85,IL90,YE95}. The instability
first sets in through $l=m$ modes. Modes with larger $l$ become unstable
at lower rotation rates but viscosity limits the interesting ones to
$l \leq5$. For an $N=1$ polytrope, the critical values of $T/W$ for the
$l=3,4$ and 5 modes are 0.079, 0.058 and 0.045 respectively and these
values become smaller for softer polytropes. 

The $l=m=2$ "bar" mode
behaves considerably different than the other modes. It's critical
$T/W$ ratio is 0.14 and it is almost independent of the polytropic
index. Since soft EOSs cannot produce models with high T/W values,
the bar mode
instability appears only for stiff Newtonian polytropes of $N \leq 0.808$
\cite{Ja64,SL96}. In addition, the viscosity driven bar mode appears
at the same critical $T/W$ ratio as the bar mode driven by
gravitational radiation (we will see later that this is no longer true
in general relativity).

     The post-Newtonian computation of neutral modes by Cutler and
Lindblom \cite{CL92,Li95}
has shown that general relativity tends to strengthen the
CFS-instability. Compared to their Newtonian counterparts, critical
angular velocity ratios $\Omega_c/\Omega_0$ (where
$\Omega_0=(3M_0/4R_0^3)^{1/2}$ and $M_0$, $R_0$ are the mass and radius
of the nonrotating star in the sequence), are lowered by as much as $10
\%$ for stars obeying the $N=1$ polytropic EOS (for which the
instability occurs only for $l=m \geq 3$ modes in the post-Newtonian
approximation).

     In full general relativity, neutral modes have been determined
for polytropic EOSs of $N \geq 1.0$ by Stergioulas and Friedman
\cite{SPHD,SF97}, using a new numerical scheme. The scheme completes
the Eulerian formalism developed by Ipser and Lindblom in the Cowling
approximation (where $\delta g_{ab}$ was neglected) \cite{IL92}, by 
finding
an appropriate gauge in which the time-independent perturbation equations 
can be solved
numerically for $\delta g_{ab}$. Because linear perturbations have a
gauge freedom, four out of ten components of $\delta g_{ab}$ are fixed
by the choice of gauge. In the Ipser and Lindblom scheme, the perturbed
Euler equations are solved analytically. A complete neutral mode solution 
of the
perturbation equations is then determined by setting the frequency in the
inertial frame equal to zero and solving six perturbed field equations 
for $\delta g_{ab}$ and the perturbed equation of
energy conservation for a scalar function $\delta U$.

    The six perturbed field equations in the gauge of Stergioulas and
Friedman are of different types. Three are second order ODEs, two are
elliptic and the other one is parabolic. Their solutions vanish at the
center, at infinity and on the axis of symmetry, while they are either
odd or even under reflection in the equatorial plane. The six
equations, although of different type, are solved simultaneously on a
two-dimensional grid, which extends to infinity by a redefinition of
the radial variable.  Solutions of the perturbed field equations are
obtained for a set of trial functions $\delta U_i$. In the Newtonian
limit, it was found that the real eigenfunctions can be expanded
accurately in terms of these trial functions \cite{IL90}. 

The remaining equation
to be satisfied, the perturbed energy conservation equation, can be
represented schematically as a linear operator $L$ on the eigenfunction
$\delta U$. Defining an inner product $< \delta U_j |L|  \delta U_i>$, 
for the set of trial functions, the perturbed energy conservation equation 
is satisfied, when 
\begin{equation}
det<\delta U_j|L|\delta U_i> = 0. \label{e:det}
\end{equation}
Using this criterion, one starts with slowly rotating configurations
and increases the angular velocity of the star until (\ref{e:det}) is
satisfied and a complete neutral mode solution is obtained.

     The determination of neutral modes for $N=1.0$, $1.5$ and $2.0$
relativistic polytropes shows that relativity significantly
strengthens the instability (which was already indicated in the 
post-Newtonian approximation). For the $N=1.0$ polytrope, the critical 
angular velocity
ratio $\Omega_c / \Omega_K$, where $\Omega_K$ is the angular velocity
at the mass-shedding limit at same central energy density, drops by as
much as $15 \%$ for the most relativistic configuration. This is a large 
decrease compared to the Newtonian values, which
significantly moves the onset of the instability away from the
mass-shedding limit and which strengthens it with respect to the damping
effect of viscosity.

     A surprising result, which was not detected in the post-Newtonian
approximation, is that the $l=m=2$ bar mode is unstable for relativistic
polytropes of index $N=1.0$. The classical Newtonian result for the
onset of the bar mode instability ($N_{crit} <0.808$) is replaced by 
\begin{equation}
                 N_{crit} <1.3,
\end{equation}
in general relativity. 

Also, in relativistic stars, the onset of the
gravitational radiation driven bar mode is different from the onset of
the viscosity driven bar mode. While in the Newtonian limit the two
bar modes occur at the same critical rotation ratio \cite{IM85}, relativity
strengthens the gravitational radiation instability, allowing softer
configurations to become unstable and suppresses the viscosity driven
instability allowing it to occur only for very stiff EOSs \cite{BFG97}. 

An independent determination of the onset of the CFS-instability in the 
relativistic Cowling approximation by Yoshida and Eriguchi \cite{YE97} 
agrees qualitatively with the conclusions in \cite{SF97}.

Morsink, Stergioulas and Blattning \cite{MSB98} extend the method
presented in \cite{SF97} to a wide range of realistic equations of state 
(which usually have a stiff high density region, corresponding to 
polytropes of index $N=0.5-0.7$) and
find that the $l=m=2$ bar mode becomes unstable for stars with
gravitational mass as low as $1.0 - 1.2M_\odot$.  For $1.4M_\odot$
neutron stars, the mode becomes unstable at $80 \% -95 \%$ of the
maximum allowed rotation rate.  For a wide range of equations of
state, the $l=m=2$ $f$-mode becomes unstable at a ratio of rotational
to gravitational energies $T/W \sim 0.08$ for $1.4 M_\odot$ stars and
$T/W \sim 0.06$ for maximum mass stars.  This is to be contrasted with
the Newtonian value of $T/W \sim 0.14$.  The empirical formula
\begin{equation}
  \left( T/W \right)_2 = 0.115 -0.048 \frac{M}{M_{\rm max}^{\rm sph}}, \label{emp}
\end{equation}
where $M_{\rm max}^{\rm sph}$ is the maximum mass for a spherical
star allowed by a given equation of state, gives the critical 
value of $T/W$ for the bar $f-$mode instability, with an accuracy of $4\%-6$\%,
 independent of the equation of state.
 
 Conservation of angular momentum and the inferred initial period
 (assuming magnetic braking) of $6-9$ms for the X-ray pulsar in the
 supernova remnant N157B \cite{Ma98}, suggests that a fraction of neutron stars
 may be born with very large rotational energies. The $f$-mode bar 
 CFS-instability thus appears as a promising source for the planned
 gravitational wave detectors \cite{LS95}. It could also play a major
 role in the rotational evolution, through the emission of
 gravitational waves, of merged binary neutron stars, if their
 post-merger angular momentum exceeds the maximum allowed to form a
 Kerr black hole \cite{BaS98}.

\subsubsection{CFS-Instability of Axial Modes}

     In nonrotating stars, axial fluid modes are degenerate at
zero-frequency but in rotating stars they have nonzero frequency and
are called $r$-modes in the Newtonian limit \cite{PP78,Sa82}. To 
$O(\Omega)$, their frequency in the inertial frame is
\begin{equation}
     \omega_i = -m\Omega \Bigl(1-\frac{2}{l(l+1)} \Bigr), \label{e:om}
\end{equation}
and modes with different radial eigenfunctions can be computed at
order $\Omega^2$ \cite{Koj97,And97}. According to (\ref{e:om}), 
$r$-modes with $m>0$ are prograde ($\omega_i<0$) with 
respect to a distant observer but
retrograde ($\omega_r = \omega_i+m\Omega >0$) in the comoving frame for
all values of the angular velocity.
Thus, $r$-modes in relativistic stars are generically unstable 
to the emission of gravitational waves via the CFS-instability, which was 
first discovered by Andersson \cite{A97}, in the case of slowly rotating,
relativistic stars. 
This result is confirmed analytically by Friedman and Morsink \cite{FM97},
who show that the canonical energy of the modes is negative.
 
Two independent computations in the Newtonian Cowling approximation,
by Andersson, Kokkotas and Schutz \cite{AKS98} and Lindblom, Owen and
Morsink \cite{LOM98} show that viscosity is not able to damp the
$r$-mode instability in rotating stars. In a temperature window of
$10^5$ K $<T< 10^{10}$ K, the growth time of the $l=m=2$ mode becomes
shorter than the shear or bulk viscosity damping time at a critical
rotation rate that is roughly one tenth the maximum allowed angular
velocity of uniformly rotating stars. The gravitational radiation is
dominated by the current quadrupole term.  These results suggest that
a rapidly rotating proto-neutron star will spin down to Crab-like
rotation rates within one year of its birth, because of the $r$-mode
instability.  The current uncertainties in the viscosity and
superfluid mutual friction damping times make this scenario also
consistent with somewhat higher initial spins, like the suggested
initial spin of $6-9$ms of the X-ray pulsar in the supernova remnant
N157B \cite{Ma98}.  Millisecond pulsars with periods less than $\sim
5$ms can then only form after the accretion-induced spin-up of old
pulsars and not in the accretion-induced collapse of a white dwarf.

The precise limit on the angular velocity of newly-born neutron stars
will depend on several factors, such as the strength of the bulk viscosity,
the cooling process, the superfluid mutual friction etc.
 In the uniform density approximation, the $r$-mode
instability can be studied analytically to $O(\Omega^2)$ in the angular
velocity of the star and the resulting expressions for the timescales, 
given in Kokkotas and Stergioulas \cite{KS98}, can be used to study
the effect of such factors on the instability. In \cite{KS98} it is
also shown that the minimum critical angular velocity for the 
onset of the $r$-mode instability is rather insensitive to the choice of
equation of state.

A first study on the issue of detectability of gravitational waves from 
the $r$-mode instability, is presented in \cite{OW98} (see section \ref{grw}),
while Andersson, Kokkotas and Stergioulas \cite{AKSt98} study the relevance of the
$r$-mode instability in limiting the spin of recycled millisecond pulsars.

\subsubsection{Effect of Viscosity on CFS-Instability}

     In the previous sections, we have discussed the growth of the 
CFS-instability driven by gravitational radiation in an 
otherwise nondissipative
star. The effect of neutron star matter viscosity on the dynamical
evolution of nonaxisymmetric perturbations can be considered
separately, when the timescale of the viscosity is much longer than
the oscillation timescale. If $\tau_{GR}$ is the computed growth rate of
the instability in the absence of viscosity and $\tau_s$, $\tau_b$ are
the timescales of shear and bulk viscosity, then the total timescale
of the perturbation is
\begin{equation}
      \frac{1}{\tau} = \frac{1}{\tau_{GR}} + \frac{1}{\tau_s} + 
                       \frac{1}{\tau_b}.
\end{equation}
Since $\tau_{GR} <0$ and $\tau_b$, $\tau_s>0$, a mode will grow only if
$\tau_{GR}$ is shorter than the viscous timescales, so that $1/\tau<0$.

     The shear and bulk viscosity are sensitive to several factors
and we give here a summary of what is known to date from Newtonian
and post-Newtonian computations:
\begin{itemize}
 \item Shear viscosity 

     In normal neutron star matter, shear viscosity is dominated by
     neutron-neutron scattering with a temperature dependence of
     $T^{-2}$ \cite{FI76} and computations in the Newtonian limit and
     post-Newtonian approximation show that the CFS-instability is
     suppressed for $T <10^6$ K - $10^7$ K \cite{IL91a, IL91b, YE95,
     Li95}. 

     If neutrons become a superfluid below a transition
     temperature $T_{s}$, then mutual friction, which is caused by the
     scattering of electrons off the cores of neutron vortices can
     completely suppress the instability for $T<T_{s}$. The superfluid
     transition temperature depends on the theoretical model for
     superfluidity and lies in the range $10^8$ K - $6 \times 10^9$ K
     \cite{Pa94}.

 \item Bulk Viscosity 
   
   In a pulsating fluid that undergoes compression and expansion, the
   weak interaction requires a relatively long time to re-establish
   equilibrium.  This creates a phase lag between density and pressure
   perturbations, which results in a large bulk viscosity \cite{Sa89}.
   The bulk viscosity due to this effect can suppress the
   CFS-instability only for temperatures for which matter has become
   transparent to neutrinos. \cite{LS95}, \cite{BFG96}.  It has been
   proposed that for $T>5 \times 10^9$K, matter will be opaque to
   neutrinos and the neutrino phase space could be blocked
   (\cite{LS95} see also \cite{BFG96}).  In this case, bulk viscosity
   will be too weak to suppress the instability, but a more detailed
   study is needed.

\end{itemize}
  In the neutrino transparent regime, the effect of bulk viscosity on
the instability depends crucially on the proton fraction $x_p$. If
$x_p$ is lower than a critical value ($\sim \frac{1}{9}$), only
modified URCA processes are allowed and bulk viscosity limits, but
does not suppress completely, the instability \cite{IL91a, IL91b,
YE95}. For most modern EOSs, however, the proton fraction is larger
than $\sim \frac{1}{9}$ at sufficiently high densities \cite{Lat91},
allowing direct URCA processes to take place. In this case, depending
on the EOS and the central density of the star, the bulk viscosity
could  almost completely suppress the CFS-instability in the neutrino
transparent regime \cite{Zd95} (but it will probably still not affect it 
for temperatures $T>5 \times 10^9$ K).

In conclusion, the available Newtonian computations indicate that the
CFS-instability in $f-$modes is effective in nascent neutron stars for
temperatures between $10^9$K and $10^{10}$K and possibly also above
$10^{10}$K if the star is opaque to neutrinos and the bulk viscosity
is weak. If direct URCA reactions do participate in the cooling
process, it appears that the instability can grow only for
temperatures for which the star is opaque to neutrinos. Since the
neutral mode computations in fully relativistic stars show that
relativity strengthens the instability, the above conclusion should
also hold in relativistic stars.

The uncertainties regarding the effect of viscosity on the CFS-instability 
in realistic
neutron stars will be greatly reduced by the construction of mode
eigenfunctions for fully relativistic, rotating stars.

\subsubsection{Viscosity-Driven Instability}

     A different type of nonaxisymmetric instability in rotating stars
is that driven by viscosity, which breaks the circulation of the fluid
\cite{RS63,Ja64}. The instability is suppressed by gravitational
radiation, so it can act only in cold neutron stars that become
rapidly rotating by accretion-induced spin-up. The instability sets in
when the frequency of an $l=-m$ mode goes through zero in the rotating
frame. In contrast to the CFS-instability, the viscosity-driven
instability is not generic in rotating stars. The $m=2$ mode becomes
unstable at a high rotation rate for very stiff stars and higher
$m$-modes become unstable at larger rotation rates.

     In Newtonian polytropes, the instability occurs only for stiff
polytropes of index $N<0.808$ \cite{Ja64,SL96}. For relativistic
models, the situation for the instability becomes worse, since
relativistic effects tend to suppress the viscosity instability
(while they strengthen the CFS-instability). According to recent
results by Bonazzola et al. \cite{BFG97}, for the
most relativistic stars, the viscosity driven bar mode can become
unstable only if $N<0.55$.  For $1.4 M_{\odot}$ stars, the instability
is present for $N<0.67$. 

These results are based on an approximate
computation of the instability in which one perturbs an axisymmetric
and stationary configuration and studies its evolution by constructing
a series of triaxial quasi-equilibrium configurations. During the
evolution only the dominant nonaxisymmetric terms are taken into
account. 

The method presented in \cite{BFG97} is an improvement
(taking into account nonaxisymmetric terms of higher
order) of an earlier method by the same authors \cite{BFG96}. Although
the method is approximate, its results indicate that the
viscosity-driven instability is likely to be absent in most
relativistic stars, unless the EOS turns out to be unexpectedly stiff.

An investigation of the viscosity-driven bar mode instability, using 
incompressible, uniformly rotating triaxial ellipsoids in the 
post-Newtonian approximation, by Shapiro and Zane \cite{SZ97}, also finds that
the relativistic effects weaken the instability.

\subsubsection{Gravitational Radiation from CFS-Instability}
\label{grw}

     The CFS-instability can limit the maximum angular velocity
of nascent neutron stars, but it is also a mechanism for the
generation of gravitational waves that could be strong enough to be
detected by the planned gravitational wave detectors. 

Lai and Shapiro
\cite{LS95} have studied the development of the $f$-mode 
instability using Newtonian
ellipsoidal rotating models \cite{LRS93,LRS94}. They consider the case
where a rapidly rotating neutron star is created in a core
collapse. After a brief dynamical phase, the protoneutron star becomes
axisymmetric but secularly unstable. The instability deforms the star
into a nonaxisymmetric configuration via the $l=2$ bar mode. Since the
star looses angular momentum via the emission of gravitational waves,
it spins-down until it becomes secularly stable. 

The frequency of the
waves sweeps downward from a few hundred Hz to zero, passing through
LIGO's ideal sensitivity band. A rough estimate of the wave amplitude shows
that, at $\sim 100$Hz, the gravitational waves from the CFS-instability 
could be detected out to the distance of 140Mpc by the advanced LIGO
detector. 
     This result is very promising, especially
since for relativistic stars the instability will be stronger than
the present Newtonian estimate.

The recently discovered CFS-instability in $r$-modes, is also an
important source of gravitational waves. Owen et al. \cite{OW98} 
model the development of the
instability and the evolution of the neutron star during its spin-down
phase. The evolution suggests that a neutron star formed in the Virgo
cluster could be detected by the advanced LIGO and VIRGO gravitational
wave detectors, with an amplitude signal-to-noise ratio that could be
as large as about 8, if near-optimal data analysis techniques are
developed.  Assuming a substantial fraction of neutron stars are born
with spin frequencies near their maximum values, the stochastic
background of gravitational waves produced by the $r$-mode radiation
from neutron star formation throughout the universe is shown to have
an energy density of about $10^{-9}$ of the cosmological closure density,
in the range 20 Hz to 1 kHz. This radiation is potentially detectable by
the advanced LIGO as well.

In newly born stars or in the post-merger objects in binary neutron
star mergers, rotating close to the Kepler limit, 
both the $f$ and $r$ modes will be unstable. Relativistic computations 
of growth times in rapidly rotating stars or even nonlinear evolutions, 
are needed to determine which mode will be strongest.

\begin{itemize}
  \item {\bf Going further.} The possible ways for neutron stars
   to emit gravitational waves and on their detectability are reviewed by
   Bonazzola, Gourgoulhon, Flanagan, Thorne and Schutz in 
   \cite{BG96,BGou,GBG96,FL98,Th96,Sc98}.
\end{itemize}

\vspace{1cm}
{\bf Acknowledgments}

I am grateful to K. Kokkotas and J. L. Friedman for critical reading of the
manuscript and for helpful suggestions. I would also like to thank the second
anonymous referee for his many useful comments. I am
indebted to the Departments of Physics of the Aristotle University of 
Thessaloniki, Greece and of the University of Wisconsin-Milwaukee, USA,
for their support during my military service. I am also grateful for the generous
hospitality of the Max-Planck-Institute for Gravitational Physics 
(Albert-Einstein-Institute) in Potsdam, Germany, where part of this 
paper was completed.


\end{document}